Rocking ratchet induced by pure magnetic potentials with broken reflection symmetry

D. Perez de Lara[1], F. J. Castaño[2], B. G. Ng[2], H. S. Körner[2], R. K. Dumas[3], E. M. Gonzalez[1], Kai Liu[3], C. A. Ross[2], Ivan K. Schuller[4] and J. L. Vicent[1]

[1] Departamento de Fisica de Materiales, Universidad Complutense, 28040 Madrid, Spain
[2] Department of Materials Science and Engineering, Massachusetts Institute of Technology, Cambridge MA 02139, USA
[3] Physics Department, University of California, Davis, California 95616, USA
[4] Physics Department, University of California-San Diego, La Jolla, California 92093, USA

Abstract
A ratchet effect (the rectification of an ac injected current) which is purely magnetic in origin has been observed in a superconducting-magnetic nanostructure hybrid. The hybrid consists of a superconducting Nb film in contact with an array of nanoscale magnetic triangles, circular rings or elliptical rings. The arrays were placed into well-defined remanent magnetic states by application of different magnetic field cycles. The stray fields from these remanent states provide a magnetic landscape which influences the motion of superconducting vortices. We examined both randomly varying landscapes from demagnetized samples, and ordered landscapes from samples at remanence after saturation in which the magnetic rings form parallel onion states containing two domain walls. The ratchet effect is absent if the rings are in the demagnetized state or if the vortices propagate parallel to the magnetic reflection symmetry axis (perpendicular to the magnetic domain walls) in the ordered onion state. On the other hand, when the vortices move perpendicular to the magnetic reflection symmetry axis in the ordered onion state (parallel to the domain walls) a clear ratchet effect is observed. This behavior differs qualitatively from that observed in samples containing arrays of triangular Ni nanostructures, which show a ratchet of structural origin.

INTRODUCTION

Ratchets are present in a wide variety of natural and artificial systems, ranging from biological motors to electronic rectifiers. They are characterized by the directional motion of out-of-equilibrium particles induced by a periodic asymmetric potential (e.g. a saw-tooth potential), without a net average driving force or temperature gradient. There are two main types of ratchets: i) flashing ratchets, in which the ratchet potential is time-dependent, as in molecular motors [1]; and ii) rocking ratchets when the ratchet potential is time-independent and the zero average driving force is periodic, as in superconducting rectifiers [2, 3]. The preparation and study of artificially prepared ratchets may shed light on the basic physics issues related to these nonequilibrium systems and may give rise to new and unusual applications [4-6].

A superconducting-magnetic hybrid is a well-defined artificial system in which a ratchet effect can be induced. These types of hybrid systems consist of a superconducting (e.g. Nb) film in contact with an array of magnetic (e.g. Ni) nanostructures. The ratchet effect manifests itself as a rectification of an a.c. current along specific geometric directions of the array. The underlying physical origin of this effect is the motion of superconducting vortices in the asymmetric periodic potential produced by the nanostructured array. In addition to the ratchet effect, these superconducting-magnetic hybrids exhibit a variety of other interesting effects such as quantum matching phenomena in which periodic matching is observed in the flux flow resistance, or bistable superconductivity which depends on the magnetic history [7-11]. A widely investigated ratchet of this type consists of a superconducting film in proximity to a square array of triangular magnetic nanostructures. For triangles arranged with one edge parallel to the x direction of the array, a ratchet effect is observed for an a.c. current along the x direction leading to vortex motion along the y direction, along which the potential is asymmetric. There is no ratchet for vortex motion along the x direction [2]. In this case the ratchet is caused by the intrinsic structural asymmetry of the triangular array. In addition, an interesting sign reversal of the superconducting ratchet is observed with the amplitude of the a.c. current, i.e., the rectified d.c. voltage changes sign with increasing amplitude of the a.c. driving force.

In the present work, ratchets of magnetic origin are investigated in a Nb/Ni system with structural symmetry. The superconducting vortices in the Nb film play the role of the nonequilibrium particle ensemble, the rocking ratchet potential is induced by an array of Ni rings, and the zero average driving forces are induced by a.c. currents injected into the hybrid samples. We show that the ratchet produced this way is purely magnetic in origin and has a different symmetry from the one produced by an array of triangular nanostructures with structural asymmetry [2]. We first present the fabrication and magnetic characterization of the arrays and the magnetotransport measurement of the hybrid. We then show that the rings provide very effective pinning sites for superconducting vortices and that they modify substantially the vortex lattice dynamics, especially the ratchet effect. We show that, for this system, the ratchet has a purely magnetic (i.e. non-structural) origin and has a different symmetry from a structurally-induced ratchet.

EXPERIMENTAL METHODS

The samples studied here consist of a superconducting Nb film in contact with arrays of magnetic Ni triangles, circular rings or elliptical rings. The Ni nanostructures were fabricated on Si (100) substrates, and then coated with the Nb layer. The preparation of the Ni triangles has been described previously [2]. The equilateral triangles were arranged in a square array with x-axis period 770 nm, triangle base length 620 nm, y-axis period 746 nm, triangle height 477 nm, and thickness 40 nm. The circular and elliptical Ni rings were prepared by electron-beam lithography in a Raith 150 electron-beam writer using polymethyl methacrylate resist and liftoff processing. Fabrication procedures for similar samples have been described elsewhere [12]. The 20 nm thick Ni film was deposited by sputtering in a system with a base pressure of $10^{-8}$ Torr. The geometries of the ring samples are listed in Table I. Two samples (A: elliptical rings, B: circular rings) were made without Nb overlayers, for MFM and magnetometry study, while two samples (C: elliptical rings, D: circular rings) were made with Nb overlayers. The dimensions of A and C, and of B and D, differed slightly. The 100 nm thick Nb films were deposited using magnetron sputtering on top of the nanostructured Ni arrays. Electrical leads were patterned using photolithography and etching techniques to form a 40µm × 40µm bridge, as shown in Figure 1.

The uncoated Ni rings were characterized magnetically by magneto-optical Kerr effect (MOKE) magnetometry on a Durham Magneto Optics NanoMOKE2 system [13] for in-plane magnetic fields. The beam was focused to a 30 µm diameter spot size, capturing the average reversal behavior of ~$10^3$ rings. With a repetition rate of 11 Hz, typically ~$10^3$ loops were averaged to obtain a single hysteresis curve. In addition to conventional major hysteresis loops, a first-order reversal curve (FORC) method [14-15] was also employed to help determine the magnetic state of the Ni nanostructures. After positively saturating the sample, the applied in-plane field was reduced to a given reversal field $H_R$ and the magnetization M was then measured back to positive saturation thereby tracing out a FORC. This process was repeated for increasing negative $H_R$ until negative saturation is reached and the major hysteresis loop is filled with FORCs (Fig. 2). The FORC distribution gives details of the switching process of the Ni rings.

Magnetic imaging was performed on samples made without the Nb overcoat by magnetic force microscopy (MFM, Digital Instruments Nanoscope), with a low-moment magnetic tip. Micromagnetic modeling of rings was performed using the two dimensional OOMMF code [16] with 2 nm × 2 nm cell size and parameters appropriate for pure Ni, with the cubic anisotropy oriented randomly in each cell.

The electrical resistivity of the hybrid system was measured using the standard 4-point probe method, in the geometry shown in Fig. 1, with a magnetic field applied perpendicular to the sample plane. In this geometry we are able to induce vortex motion in the x and y in-plane directions of the array. The relevant transport properties of all hybrid samples are similar, i.e. the critical temperatures are 8.10 K (elliptical ring sample), 8.15 K (circular rings) and 8.30 K (triangles), and the mean free paths are $8 \times 10^{-10}$ m (elliptical rings), $12 \times 10^{-10}$ m (circular rings) and $10 \times 10^{-10}$ m (triangles).

RESULTS AND DISCUSSION

The magnetic state of the ring arrays was examined for three cases: circular rings with the in-plane field along one axis of the square array, elliptical rings with the field along the long axis (LA) of the ellipses, and elliptical rings with the field along the short axis (SA) of the ellipses. In all three cases the FORC magnetometry data (Fig. 2) and the MFM data (Fig. 3) indicate that the rings form onion states at remanence after saturation at fields > 600 Oe.
Onion states consist of two domains separated by domain walls at opposite sides of the ring, aligned parallel to the direction of the saturating field. For rings of these dimensions, thickness, and material, the simulation shows that the domain walls have transverse character with in-plane magnetization. These results are in agreement with previous results on thin film magnetic rings [18, 19]. For the circular rings and the elliptical rings magnetized along their long axis, practically all of the rings form onion states of identical orientation, and the dark or bright contrast of the domain walls is evident in Figs. 3g and 3h respectively. For the elliptical rings magnetized along the short axis, some of the rings (<10%) form 'vortex' states with no domain walls and no magnetic contrast, as seen in Fig. 3i, but the majority form parallel onion states with domain walls located along the minor axis of the ellipse.

In contrast, demagnetization of the ring arrays is expected to lead to a magnetic configuration consisting of onion states oriented in different directions, in addition to 'vortex' states. In this case the stray magnetic field fluctuates aperiodically across the sample providing a random magnetic potential landscape.

The magnetic properties of the array of Ni triangles have been reported previously [17]. At remanence after in-plane saturation the triangles form magnetic vortex states exhibiting both clockwise and counterclockwise chiralities and 'up' and 'down' polarities of the vortex core.

In general, structural defects are very effective vortex pinning centers [20]. In these types of systems close to the superconducting critical temperature, the magnetoresistance exhibits pronounced periodic minima when the vortex density is an integer multiple of the pinning site density [7]. Hence, the number of vortices n per Ni nanostructure is controlled by the external magnetic field $H_z$ perpendicular to the sample, and can be characterized using magnetoresistance measurements. Since the perpendicular applied magnetic fields are less than 400 Oe, we expect that the remanent states of the rings are not substantially modified during the experiment due to the high demagnetizing factors of the thin film structures. Fig 4 shows magnetoresistance vs. $H_z$ of three hybrid samples that were initially in the demagnetized state, i. e. samples as deposited. The triangular, elliptical rings and circular rings have similar geometric sizes and periodicities, and the three samples show similar periodicity in the magnetoresistance fluctuations. However, the samples containing rings exhibit a significantly larger number of minima than the ones containing triangular nanomagnets. Thus the rings in these hybrids are very effective pinning centers and are able to pin a larger number of vortices than hybrid samples containing magnetic triangles.

Fig. 5 shows a comparison between the ratchet effect in samples containing triangular and elliptical rings measured at a perpendicular field $H_z$ corresponding to n=3, i.e. three vortices per pinning site. The a.c. injected current is parallel to the triangular base or the ellipse short axis. Therefore, the vortex motion is from triangle base to tip, or parallel to the ellipse major axis, respectively. The motion is caused by the Lorentz force, $\vec{F}_L = \vec{J} \times \vec{z}\phi_0$ (where $\phi_0$ is the quantum fluxoid, $\vec{z}$ is a unit vector parallel to the applied magnetic field and J is the a.c. current density) acting on the vortices. Although the time-averaged force on the vortices $\langle F_L \rangle = 0$, a non-zero d.c. voltage drop is observed for the triangles, as seen previously [2], and for the rings measured at remanence after saturation, where they are in parallel onion states. Since the electric field $\vec{E} = \vec{B} \times \vec{v}$ ($\vec{v}$ and $\vec{B}$ being the vortex-lattice velocity and the applied magnetic induction, respectively), the voltage drop ($V_{dc}$) along the direction of the injected current probes the vortex motion along the perpendicular direction. Therefore, the time-averaged vortex velocity is $\langle v \rangle = V_{dc}/dB$, where $d$ is the distance between the voltage contacts.

As noted above, the Ni triangles in the demagnetized state develop a magnetic vortex state, with random chirality (clockwise or counter clockwise) and polarity (up or down) [13, 17]. Thus, in the absence of a periodic magnetic potential, the ratchet effect in this case must originate from the structural shape asymmetry of the array of triangles [2]. It is interesting to note that for the purely geometric structural asymmetry exhibited by these triangles the ratchet effect is odd with the direction of the applied magnetic field, i.e. reversing the magnetic field reverses the sign of $V_{dc}$.

On the other hand, the samples with elliptical rings are structurally symmetrical and therefore show no ratchet effect in the demagnetized state (empty circles in Fig. 5), but an unambiguous ratchet effect (solid black circles in Fig. 5) occurs when the elliptical rings are magnetized in parallel onion states. The latter ratchet is attributed to a pure magnetic asymmetry. This magnetic origin is further confirmed by the fact that the sign of $V_{dc}$ does not change when the applied magnetic field direction is reversed, i.e., the ratchet effect is even. The magnetic ratchet effect (in the rings) and the shape-induced ratchet effect (triangles) share certain similarities: i) both are adiabatic, i.e., independent of frequency [2] up to the highest attainable frequency in our experiments (10 kHz), and ii) both show the same experimental trends, i.e., the magnitude of the output d.c. voltage increases (at constant applied field) with decreasing temperature and (at constant temperature) with increasing applied magnetic field [21]. There is also an important difference between the two ratchet effects: for the ring samples the sign of $V_{dc}$ (the ratchet polarity) does not change with increasing number of vortices or with the a.c. input driving currents, i.e., there is no vortex ratchet reversal for a purely magnetically-induced ratchet. In contrast, structurally asymmetric ratchets do exhibit a sign reversal [2, 3, 21] in which the d.c. output voltage polarity can be tuned by external parameters. For instance, at n = 4 the sign of the d.c. voltage changes with increasing amplitude of the a.c. drive current as shown in the inset of Fig. 5.

Fig. 6 shows the ratchet effect for both circular and elliptical rings magnetized in parallel onion states for different temperatures and vortex motion directions (along or perpendicular to the magnetization axis, i.e. parallel or perpendicular to the domain walls). When the vortex motion is perpendicular to the domain walls (and to the direction of the initial saturating in-plane field) the ratchet is absent, as expected from a purely magnetic effect, because the magnetic potential is symmetrical. The driving force threshold and the magnitude of the ratchet effect increase with decreasing temperature as found earlier [21]. The ratchet in circular rings shows similar trends to the elliptical rings. The magnitude of $V_{dc}$ for the circular sample is similar to that measured for the elliptical samples for vortex motion along the minor axis, but much smaller than the $V_{dc}$ measured for elliptical rings when the vortex motion is along the major axis. This result suggests that the separation between the magnetic potential wells plays an important role in the strength of the ratchet effect and therefore in the amplitude of the d.c. output voltage.

As we have seen, the magnetization direction in the ring dictates the direction along which the vortices have to move to produce a ratchet effect. This is analogous to the triangular elements, in which the structural asymmetry defines the directions along which a ratchet effect can be observed [2, 21]. However in the case of the rings, the ratchet is tuneable according to the magnetic state of the pinning sites, and can be 'switched off' by demagnetizing the array. The results shown in Fig. 6 indicate that an asymmetric magnetic potential is only found when the vortices move along the magnetization direction parallel to the direction of the initial saturating field and to the remanent domain walls. Experimentally, the asymmetric ratchet potential is therefore clearly related to the differences in the interaction between the vortex motion and the magnetization direction in the rings. The variation of the amplitude of the ratchet effect for different directions and types of ring arrays implies that the interaction of the vortex motion with the magnetic fields produced by the onion states plays an important role in the magnitude of the effect.

Recently, a pure magnetic ratchet potential has been discussed theoretically [22, 23]. This ratchet is based on pinning by arrays of magnetic dipoles, and this model has been applied in the case of superconducting vortices. In this theory, the origin of this pure magnetic ratchet effect is the periodic vortex pinning potential with broken reflection symmetry created by the magnetic dipoles. This pinning of the vortex lattice by dipole arrays occurs if the vortex lattice periodicity is pinned by and commensurate with the dipole array periodicity. This theoretical model is applicable here, since the remanent onion state induces periodic dipolar magnetic potentials, and the applied field $H_z$ was chosen such that the vortex lattice is commensurate with the periodicity of the magnetic rings, based on the magnetotransport measurements.

Superficially similar results [24] in hybrids of Pb films with array of triangular Co microrings and Al films with Co microbars have been reported. These results were analysed with the aforementioned model [22, 21] invoking the appearance of vortex-antivortex pairs. However, contrary to our results, several polarity reversals were observed with increasing number of vortices. Furthermore, a non-zero ratchet effect was found even in zero applied fields. Since some of these experiments rely on arrays which are structurally asymmetric, the reported ratchet effects may not be purely magnetic in origin, and the relationship to the results reported here is not yet clear.

CONCLUSIONS

     Superconducting-magnetic hybrid systems based on periodic arrays of Ni rings in contact with Nb films show rocking ratchet effects when the superconducting vortex lattice moves in an asymmetrical magnetic potential landscape induced by the periodic array. The most relevant results can be summarized as follows:

i)   The ratchet effect is observed only when the ring array is magnetized at remanence into parallel onion states, and not when the array is demagnetized.

ii)  The magnetic ratchet observed here resembles the ratchet produced by structurally asymmetric pinning sites in that both are adiabatic, and the amplitude and threshold of the driving force increases with decreasing temperature. However, they differ qualitatively in that the magnetic ratchet does not change its polarity when the applied magnetic fields are reversed or the driving force strength (a.c. current) is increased.

iii) The origin of the ratchet effect lies in the interaction between the superconducting vortex screening currents and the periodic asymmetric stray magnetic fields produced by the rings when placed in parallel remanent onion states. This interaction establishes a net direction for the vortex motion when it is driven by a zero average a.c. current. While the detailed mechanism for this interaction has not been established, our experiments show that this effect must be purely magnetic in origin.

This work has been supported by Spanish Ministerio Ciencia e Innovacion, grants NAN2004-09087, FIS2005-07392, Consolider CSD2007-00010, FIS2008-06249 (Grupo Consolidado) and CAM grant S-0505/ESP/0337, Fondo Social Europeo, CITRIS, the US National Science Foundation, and the INDEX program of the Nanoelectronics Research Institute. IKS acknowledges support from UCM and Banco Santander during his sabbatical stay in Universidad Complutense at Madrid, and K.L. acknowledges a UCD Chancellor's Fellowship.

Figure Captions

Fig. 1. Optical image of the cross-shaped bridge (40 µm x 40 µm) and electron micrographs of the arrays of elliptical and circular rings at higher magnifications.

Fig. 2. Families of FORC's for an array of elliptical Ni rings (sample A in Table I) where the applied field is (a) parallel and (b) perpendicular to the long axis (major axis) of the ellipses. (c) FORC's measured on an array of circular Ni rings (sample B in Table I). The insets show a sketch of the onion magnetic states and the asymmetric magnetic reflection axis (double arrows).

Fig. 3. (a,b,c) OOMMF simulation of the remanent state of (a) a circular ring after 5000 Oe saturation, (b) an elliptical ring after 5000 Oe saturation along the long axis, (c) an elliptical ring after 5000 Oe saturation along the short axis; (d,e,f) topographic and (g,h,i) magnetic remanent images of (d,g) the circular ring array at 65nm lift height; (e,h) the elliptical ring array at 50 nm lift height, long axis field; (f,i) the elliptical array at 65 nm lift height, short axis field. Images are 6 µm square. The white dotted lines indicate the locations of some of the rings in the MFM images.

Fig. 4. Magnetoresistance of hybrid Nb film (100 nm thickness) on top of arrays of Ni triangles (full triangles), elliptical rings (solid dots, sample C, see Table I) and circular rings (open dots, sample D) measured with the magnetic nanostructure arrays initially in the demagnetized state. The temperature $T/T_c$ was 0.99. Current densities: $J = 8.0 \times 10^2$ $Acm^{-2}$ (triangle sample); $J = 2.5 \times 10^3$ $Acm^{-2}$ (elliptical rings); $J = 6.0 \times 10^3$ $Acm^{-2}$ (circular rings). The data for the triangles has been scaled by a factor of two for clarity.

Fig. 5. Measurements of vortex ratchet effects in Nb/Ni hybrids with triangular and elliptical ring Ni nanomagnet arrays. The Ni triangles were in the demagnetized state. The Ni rings were measured in both the demagnetized state (open circles) and in parallel onion states after saturation along the long axis (solid circles). The vortex lattice motion corresponds to translation from the triangle base to the apex in the array of triangles and along the major axis in the case of the array of elliptical rings. The frequency of the injected a.c. current is 10 kHz and $T/T_c = 0.98$. The inset shows ratchet reversal in the array of Ni triangles obtained with increasing a.c. drive current.

Fig. 6. Experimental data showing the ratchet effect for samples with elliptical and circular Ni rings for different vortex motion directions and temperatures. In all cases the rings were magnetized into parallel onion states. The vortex motion is along either the major axis (LA) or minor axis (SA) in the case of the elliptical rings. All the measurements were made with the vortex motion along the direction of the domain walls (DW), parallel to the initial in-plane saturating field, except for the data shown with solid circles in which there is no ratchet effect. The measurement temperatures were 0.98 $T_c$ (solid triangles and solid diamonds) and 0.97 $T_c$ (open triangles and open diamonds). $F_L$ is the Lorentz force on the vortices and <v> their average velocity.

Table I
Dimensions of the circular and elliptical ring arrays in the Ni/Nb hybrid samples.

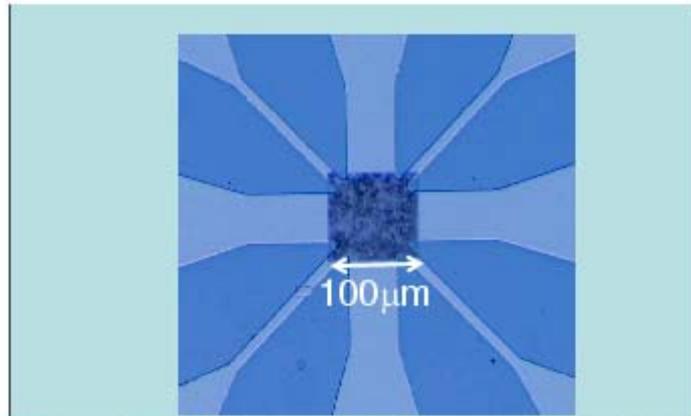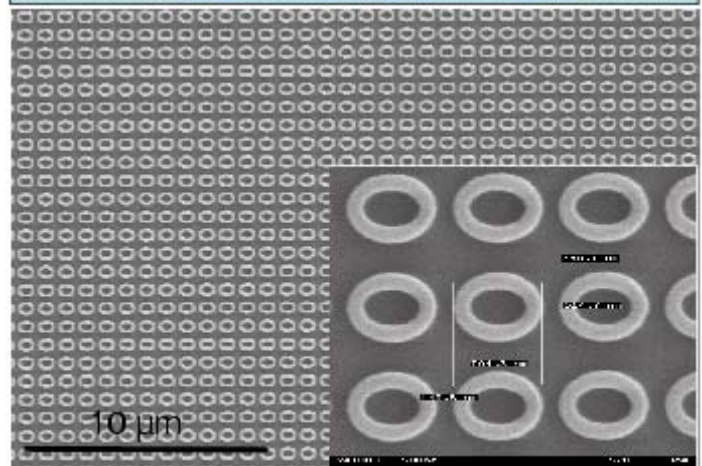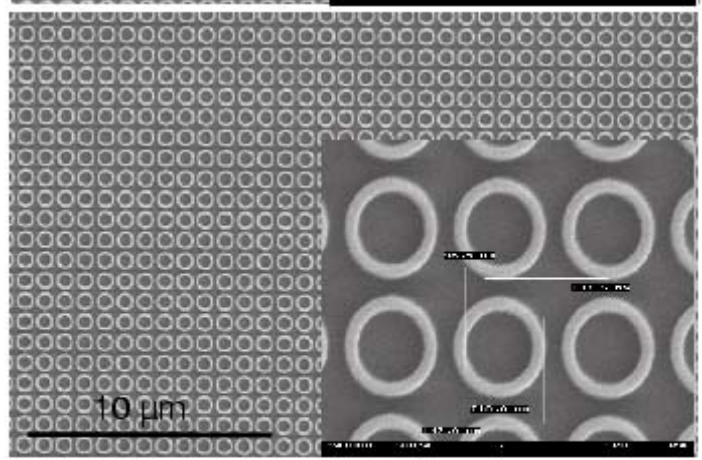

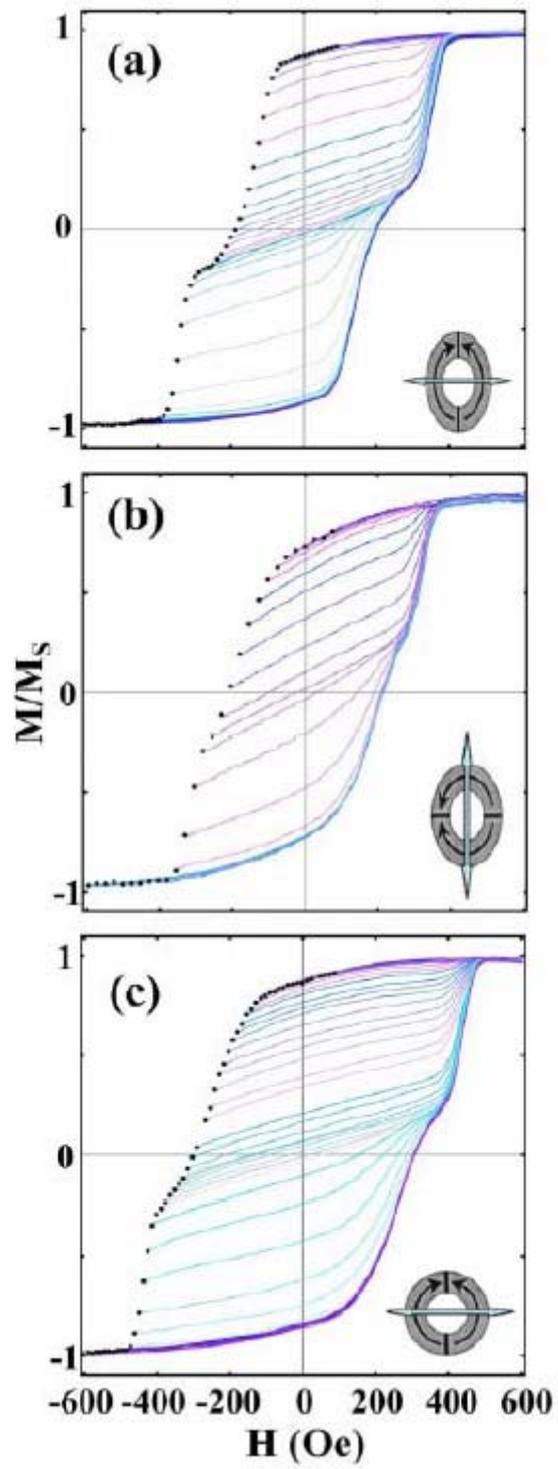

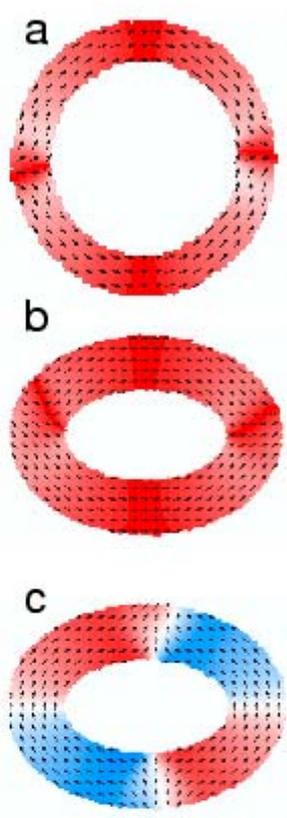
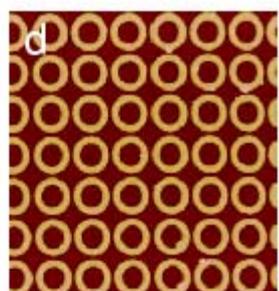
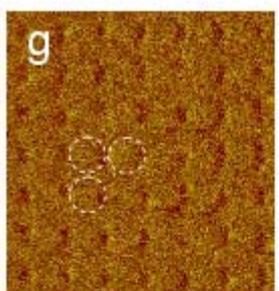
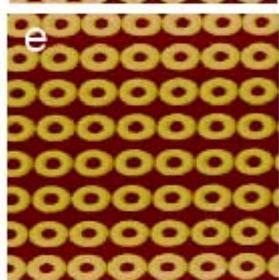
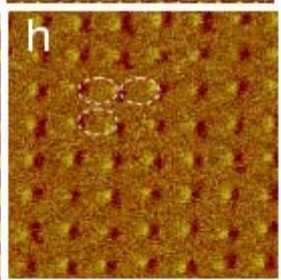
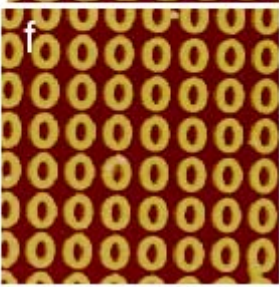
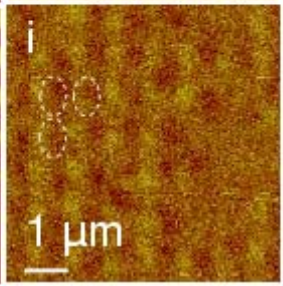

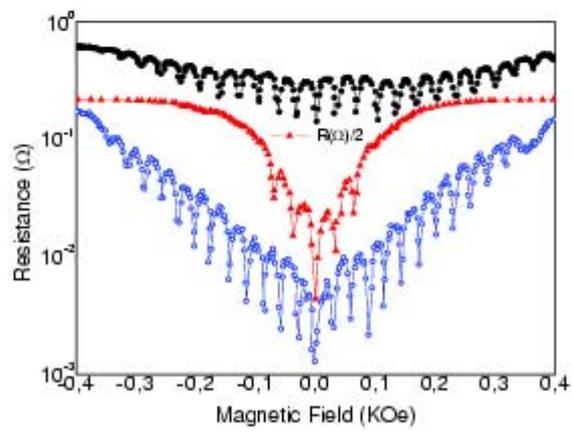

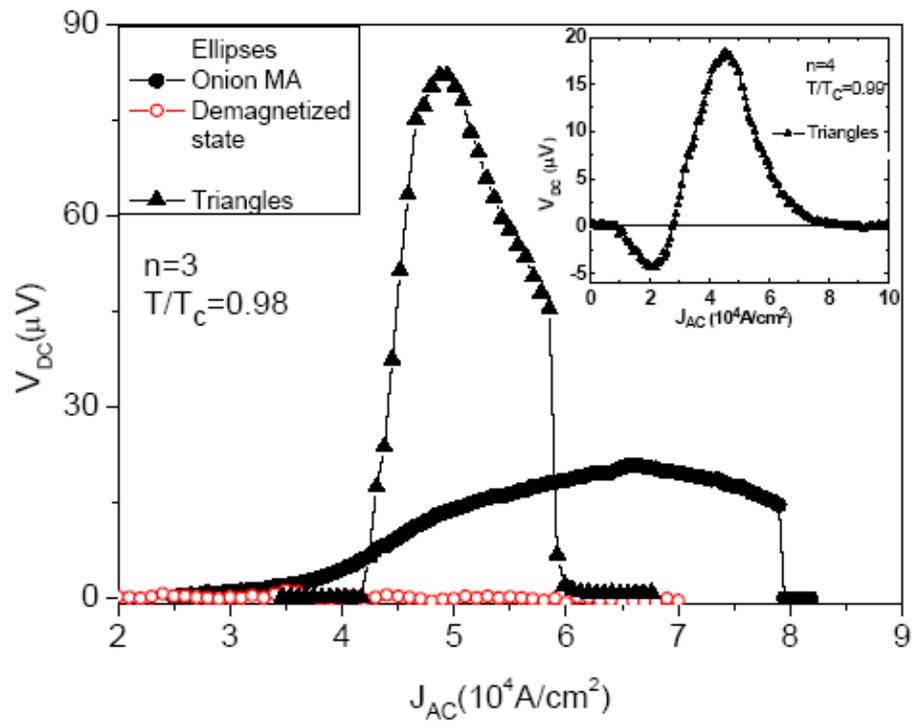

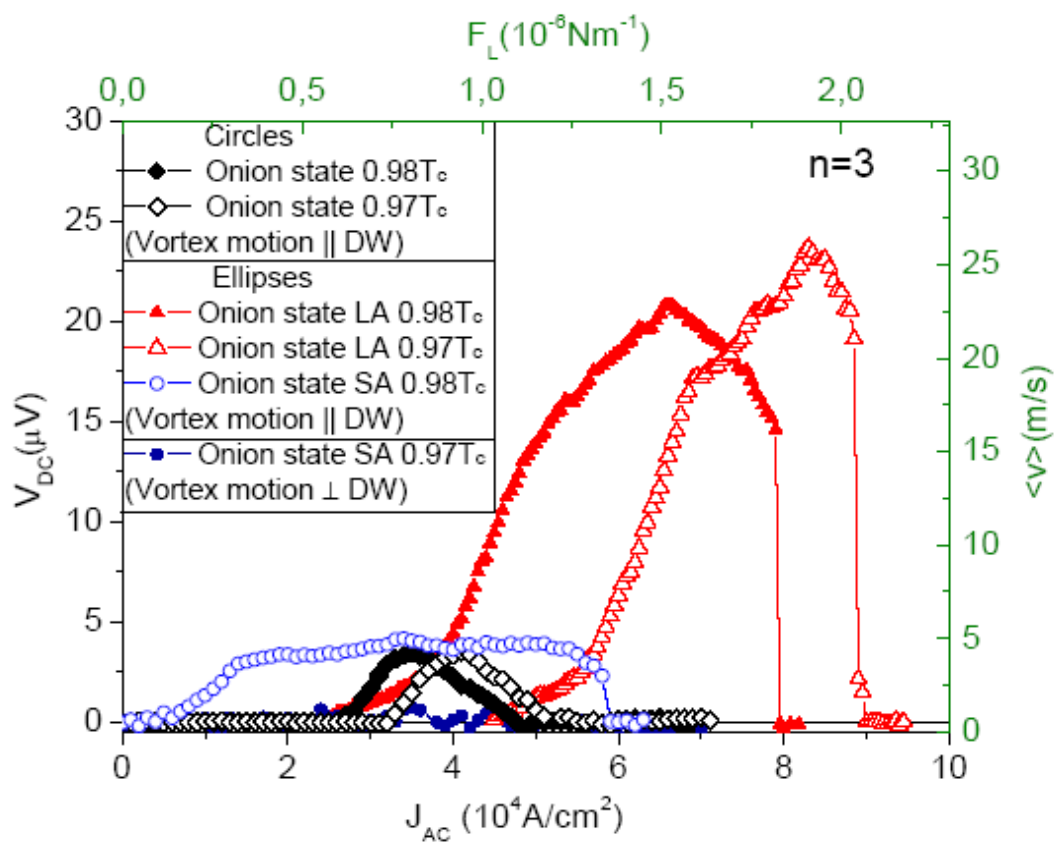

| Rings | Length (nm) | | Period (nm) | | Spacing (nm) | | Line Width (nm) | Ni \| Nb Thickness (nm) | Area ($\mu m^2$) |
|---|---|---|---|---|---|---|---|---|---|
| | Long | Short | Long | Short | Long | Short | | | |
| A ELLIPSES | 700 | 520 | 820 | 730 | 120 | 210 | 140 | 20 \| 0 | 100x100 |
| B CIRCLES | 700 | | 820 | | 120 | | 100 | 20 \| 0 | 100x100 |
| C ELLIPSES | Long | Short | Long | Short | Long | Short | 130 | 20 \| 100 | 100x100 |
| | 725 | 560 | 820 | 745 | 95 | 185 | | | |
| D CIRCLES | 700 | | 820 | | 120 | | 125 | 20 \| 100 | 100x100 |